\documentclass[12pt]{article}
\usepackage[a4paper, total={6in, 10in}]{geometry}

\usepackage{graphicx}

\usepackage{booktabs} 
\usepackage{multirow} 
\usepackage{amsmath} 
\usepackage{wrapfig} 
\usepackage{svg}
\usepackage{amsthm} 
\theoremstyle{definition} 
\usepackage{soul}
\usepackage{makecell}
\usepackage{subcaption}
\usepackage{caption}
\usepackage[table]{xcolor}
\usepackage{mathtools}
\usepackage{tcolorbox}

\usepackage{pifont} 

\usepackage{tikz}
\usetikzlibrary{chains,shapes.multipart}
\usetikzlibrary{shapes}
\usetikzlibrary{automata,positioning}
\usetikzlibrary{calc}
\usepackage{tikzscale}
\usepackage{makecell}
\usepackage{tabularx}
\usepackage{tkz-graph}
\usepackage{float}
\usepackage{amsfonts}
\usetikzlibrary{shapes.geometric}
\usepackage{amssymb}
\usepackage{braket}
\tcbset{
    sharp corners,
    colback = white,
    before skip = 0.5cm,    
    after skip = 0.5cm      
}  
\definecolor{main}{RGB}{0,0,0}
\newtcolorbox{boxA}{
    fontupper = \bf\color{main}, 
    boxrule = 1.5pt,
    colframe = main,
    rounded corners,
    arc = 5pt   
}

\newtheorem{stmt}{Statement}
\usepackage{hyperref} %

\title{Qubit-Efficient QUBO Formulation for Constrained Optimization Problems}
\author{$\text{Meerzhan Kanatbekova}^{1}$, $\text{Vincenzo De Maio}^{1,2}$, $\text{Ivona Brandić}^{1}$}
\date{$^1$ Institute of Computer Engineering, TU Wien, Austria, \\ $^2$ School of Computing and Mathematical Sciences, University of Leicester, UK}

\begin{document}


\maketitle




\begin{abstract}
    Quantum computing has emerged as a promising alternative for solving combinatorial optimization problems. The standard approach for encoding optimization problems on quantum processing units (QPUs) involves transforming them into their Quadratic Unconstrained Binary Optimization (QUBO) representation. However, encoding constraints of optimization problems, particularly inequality constraints, into QUBO requires additional variables, which results in more qubits. Considering the limited availability of qubits in NISQ machines, existing encoding methods fail to scale due to their reliance on large numbers of qubits.

    We propose a generalized exponential penalty framework for QUBO inequality constraints inspired by a \textit{class of exponential functions}, which we call exponential penalization. 
    This paper presents an encoding strategy for inequality constraints in combinatorial optimization problems, inspired by a \textit{class of exponential functions}, which we call exponential penalization. The initial idea of using exponential penalties for QUBO formulation was introduced by Monta\~nez-Barrera et al. by applying a specific exponential function to reduce qubit requirements. 
    In this work, we extend that approach by conducting a comprehensive study on a broader class of exponential functions, analyzing their theoretical properties and empirical performance.  
    Our experimental results demonstrate that \textit{an exponential penalization} achieves 57\%, 83\% qubit number reduction for Bin Packing Problem (BPP) and Traveling Salesman Problem (TSP), respectively. And we demonstrate comparable solution quality to classical with a probability of 6\% and 21\% accuracy for BPP with 8 and TSP with 12 qubits, respectively.

\end{abstract}

\section{Introduction}

Optimization problems play a critical role in diverse industrial and scientific domains, yet their increasing scale has led to computational demands that exceed the capabilities of classical methods. As transistor miniaturization slows and Moores Law approaches its physical limits, classical computing architectures face inherent barriers in efficiently addressing large-scale combinatorial problems. These challenges have directed growing attention toward quantum computing, which has recently become an active field of research for solving complex optimization problems~\cite{hervert2020production, qi2020optimization, gupta2022enhanced, sankar2024benchmarking, abbas2024challenges}.

To solve the optimization problem on a quantum computer, first, it has to be encoded in quantum states. The standard approach for encoding optimization problems on quantum processing units (QPUs) involves transforming them into their Quadratic Unconstrained Binary Optimization (QUBO) representation~\cite{glover2022quantum}. Within the QUBO framework, the constraints of the optimization problem are incorporated into the objective function as penalty terms.

Existing constraint-encoding methods, such as linear, quadratic penalties~\cite{dantzig1990origins}, have shown effectiveness in classical optimization frameworks but often lack scalability when applied to quantum systems~\cite{mirkarimi2024quantum}, since inequality constraints require additional decision variables in their QUBO formulation. The commonly used approach involves incorporating extra variables (so-called slack variables) to encode these constraints~\cite{dantzig1990origins}. However, this method leads to an increase in the dimensionality of the problem, which poses a significant challenge to quantum computers, where the number of qubits is limited.

To address these challenges, exponential penalization for inequality constraints has been proposed in~\cite{montanez2024unbalanced}. Exponential penalties offer a steeper growth rate compared to quadratic penalties, allowing for more efficient enforcement of constraints with potentially lower qubit count. The authors in~\cite{montanez2024unbalanced} demonstrated that exponential penalization can substantially reduce the number of qubits required for problems such as the Bin Packing Problem (BPP), the Travelling Salesman Problem (TSP), and the Knapsack Problem (KP). Quadratic penalties grow polynomially, providing moderate enforcement that can be too forgiving in some cases. In contrast, exponential penalties grow rapidly with the magnitude of constraint violations, allowing them to penalize large violations more severely and potentially achieve better constraint satisfaction with fewer qubits. 

However, the work in~\cite{montanez2024unbalanced} was limited to a single type of exponential function. In this work, we extend the concept of exponential penalization by considering a \textit{class of exponential penalty} functions. This class captures various exponential growth behaviors that can be tuned to enforce constraints more effectively across different problem instances. Specifically, we investigate three different forms of exponential penalty functions derived from the main formulation and analyze their impact on solution quality, time complexity, and convergence behavior. Our focus is on combinatorial optimization problems, including BPP and TSP, as representative benchmarks.

\begin{figure}
    \centering
\includegraphics[width=0.9\linewidth]{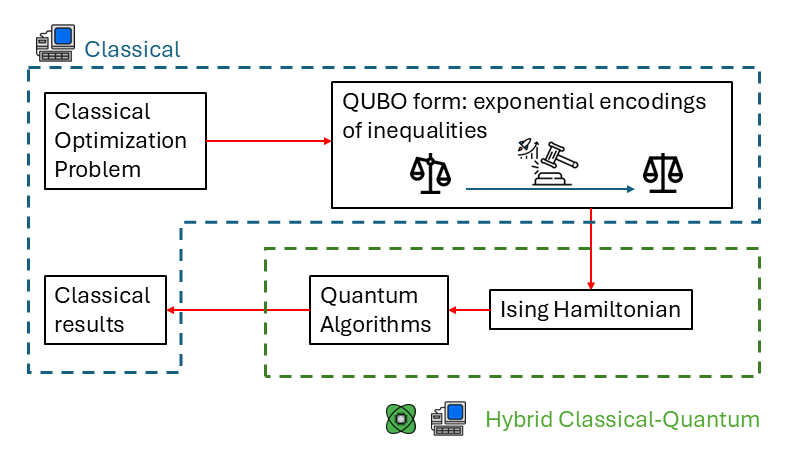}
    \caption{Overview of the optimization problem solution pipeline on classical-quantum computers, featuring the QUBO formulation derived using exponential penalty functions.}
    \label{fig:overview}
\end{figure}

While the detailed methodology and experimental setup are provided in subsequent sections, an overview of the computational pipeline is illustrated in Figure~\ref{fig:overview}. Given optimization problems such as the BPP and the TSP, we begin by applying our proposed exponential encoding method to derive the QUBO formulation. This efficient QUBO representation is then transformed into an equivalent Ising Hamiltonian, which serves as the input to quantum computing frameworks. We subsequently solve the problem on a quantum computer (or a quantum simulator) using the Quantum Approximate Optimization Algorithm (QAOA) and obtain the final solution. The experimental framework utilizes IBM's Qiskit simulator, enabling controlled evaluation of solution quality, qubit requirements, and execution times under realistic quantum settings.

In summary, the \textbf{key contributions} of the paper are:
\begin{itemize} \renewcommand\labelitemi{--}
    \item We propose an \textit{exponential penalization method} using \textit{a class of exponential penalty functions} to encode inequality constraints in optimization problems. We also provide different functions within the class of exponential functions. 
    \item We empirically evaluate the proposed method on generated datasets and compare it to existing baseline approaches with respect to execution time, solution quality, qubit count compression/reduction rate.
    \item Our extensive experiments on IBM simulator demonstrate that \textit{an exponential penalization} achieves 57\%, 83\% qubit count reduction rate while preserving 6\%, 21\% solution accuracy for BPP, TSP respectively. 
    \item We optimize the exponential penalty parameters such as $r,s,p$ and provide an empirical study on them by determining the significance factor of each. 
\end{itemize}

The structure of the paper is as follows. Section~\ref{sec:related_work} reviews existing encoding techniques and their relevance to quantum optimization. Section~\ref{sec:background} describes the background and introduces the benchmark problems. Section~\ref{sec:proposed_method} introduces the proposed exponential penalization, including a detailed theoretical analysis. Section~\ref{sec:experiment} describes the experimental setup and metrics. Section~\ref{sec:results} presents experimental results demonstrating the effectiveness of the approach in both classical and quantum optimization settings. Finally, Section~\ref{sec:conclusion} concludes the paper and outlines potential directions for future research.

\section{Related Work} \label{sec:related_work}
Constraint handling in optimization problems has been extensively studied, both in classical and quantum computing domains. In the literature, two primary penalization strategies are commonly used: interior and exterior penalty methods~\cite{coello2022constraint}. Interior penalty methods penalize violations within the feasible region, while exterior penalty methods penalize infeasible solutions. Exterior penalties typically include linear and quadratic, each with distinct enforcement characteristics.

Several recent works have proposed more qubit-efficient encodings of optimization problems. Tan et al.~\cite{Tan2021qubitefficient} tackle optimization problems by proposing an encoding scheme for QUBO models with $n$ variables that can be implemented on $O(log n)$ number of qubits. They propose a strategy using $n$ ancilla qubits and $r$ register qubits to divide the QUBO problem into $2r$ subsystems of $n$ classical variables, requiring a total of $t = n + r$ qubits. For example, optimization problems with a problem size of $10^4$ can be solved only using no more than 15 qubits. However, this method focuses primarily on encoding the final QUBO representation more efficiently, and does not directly address constraint handling or inequality encoding.

A specialized encoding for the Travelling Salesman Problem (TSP) was proposed in~\cite{10646502} through permutation encoding based on Heap's algorithm. This approach maps each TSP route to a unique permutation index, significantly reducing the qubit requirement to $\log(n!)$ compared to the $n^2$ qubits typically needed in standard QUBO formulations. While this method is highly qubit-efficient for routing problems, it is not directly applicable to general constrained optimization problems, especially those with complex inequality constraints.

Leonidas et al.~\cite{leonidas2024qubit} investigated encoding strategies for the Vehicle Routing Problem with Time Windows (VRPTW). They compared minimal and full encoding schemes across problem instances ranging from 11 to 3964 routes, evaluating performance on both simulators and quantum hardware. Their results showed that minimal encoding could reduce qubit usage, but each classical variable still typically required dedicated qubits unless further optimizations were introduced. Their focus was primarily on mapping efficiency rather than constraint flexibility.

Roch et al.~\cite{roch2023effect} empirically studied the effect of penalty weight selection on the minimum spectral gap of constrained optimization problems when solved via quantum annealing. They demonstrated that carefully tuned penalty weights can improve solution quality by enlarging the spectral gap, and they proposed using machine learning models to predict optimal penalty coefficients. However, their work is focused on penalty magnitude selection rather than exploring different penalty functions.

Montanez et al.~\cite{montanez2024unbalanced} proposed an alternative method for encoding inequality constraints without introducing slack variables. Their approach uses an exponential penalty function that rapidly penalizes constraint violations, reducing the number of qubits needed compared to traditional methods. However, their study is limited to a single form of exponential function and does not investigate how different exponential growth rates or types of exponential functions could affect performance, solution quality, or qubit efficiency.

In contrast to these works, our study proposes a generalized exponential penalization framework that introduces a tunable class of exponential functions for inequality constraint encoding. Unlike previous methods, our approach:
\begin{itemize}
    \item Systematically explores multiple exponential penalty functions.
    \item Optimizes the penalty parameters empirically for each problem class.
    \item Demonstrates substantial qubit savings without sacrificing solution quality.
    \item Provides flexibility across diverse combinatorial optimization problems.
\end{itemize}

\section{Background} \label{sec:background}

\subsection{Combinatorial Optimization Problems}
A combinatorial optimization problem (COP) is a set of components $(S, S', f, g, h)$, where $S$ is the finite search space, $S'\subset S$ is the set of feasible solutions in $S$, $f:S'\rightarrow \mathbb{R}$ is an objective function. The goal is to find $x^*\in S'$ that maximizes or minimizes the objective function $f(x)$, subject to constraints $g$ and $h$ defined on $S$. COPs are defined as: 

\begin{align}
\min/\max \quad & f(x) \label{eq:cop1} \\
\text{s.t.} \quad 
& g_i(x) = 0, && \forall i \label{eq:cop_eq_const} \\
& h_j(x) \le 0, && \forall j \label{eq:cop_ineq_const}
\end{align} \label{eq:cop}

where $f:S'\rightarrow \mathbb{R}$ is an objective function, $g_i:S'\rightarrow \mathbb{R}, i=1,...,m$ are equality constraints and $h_j:S'\rightarrow \mathbb{R}, j=1,...,k$ are inequality constraints. In contrast to continuous optimization problems, where the solution space is continuous, in COP the feasible solutions are discrete, i.e., the focus is on discrete structures, such as graphs or permutations. Moreover, certain COPs are known to be NP-hard, implying that the solution is computationally challenging and may demand superpolynomial time in worst-case scenarios.  In this work, we experimented proposed encoding method on two well-known COPs: Bin Packing Problem (BPP) and Travelling Salesperson Problem (TSP).

\subsubsection{The Bin Packing Problem (BPP)} is a well-known combinatorial optimization problem that arises in logistics, manufacturing, and resource allocation~\cite{v2023hybrid, MANDAL199891}. Given a set of items with known weights, the goal is to pack them into a minimum number of bins while ensuring that the total weight in each bin does not exceed a fixed capacity.

We consider the \emph{1-d offline BPP}. Each item $x_{ij}$ has a weight, $w_i$, and has to be assigned to a bin $B_j$ using a minimum number of bins with fixed capacity $C$. Formally,

\begin{align}
\min \quad & \sum_{j=1}^{K} B_j \label{eqn:bpp} \\
\text{s.t.} \quad & \sum_{j=1}^{K} x_{ij} = 1, && \forall i \label{eqn:bpp_constraint_1} \\
& \sum_{i=1}^N w_{i}x_{ij} \le C \cdot B_j, && \forall j \label{eqn:bpp_constraint_2}
\end{align}

where $N$ and $K$ are the number of items and bins, respectively.  The equality constraint $g(x_{ij})$ in \ref{eqn:bpp_constraint_1} implies that each item can be packed into at most one bin, and the inequality constraint $h(x_{ij})$ in \ref{eqn:bpp_constraint_2} says that the weight of bin item can not exceed the given capacity $C$. 
\begin{table}[h!]
\centering
\caption{Summary of notation and descriptions for the Bin Packing Problem (BPP) with one type of example solved as part of the experiments. }
\label{tab:bpp_notation}
\begin{tabular}{|c|c|c|}
\hline
\textbf{Notation} & \textbf{Description} & \textbf{Example} \\ \hline
$N$ & Total number of items & 3 \\ \hline
$K$ & Total number of bins & 2 \\ \hline
$w_i$ & Weight of item $i$ & [25,25,30]\\ \hline
$C$ & Capacity of a single bin & 100 \\ \hline
$B_j$ & Binary variable: $1$ if bin $j$ is used, $0$ otherwise & \{0,1\}\\ \hline
$x_{ij}$ & Binary variable: $1$ if item $i$ is assigned to bin $j$, $0$ otherwise & \{0,1\}\\ \hline
\end{tabular}
\end{table}
In BPP, determining whether a given set of items can be packed into at most $K$ bins is NP-complete, while the optimization part, which seeks to minimize the number of bins used, is NP-hard. This complexity arises from the difficulty of partitioning items optimally among bins, which is equivalent to solving a subset sum problem, a known NP-complete problem. Additionally, the number of feasible packing assignments grows exponentially as $O(K^N)$, making it computationally infeasible for large instances. The Table~\ref{tab:bpp_notation} summarizes the notations used for BPP formulation. 

\subsubsection{Traveling Salesman Problem (TSP)}
Let $G=(V,E, w: E \mapsto \mathbb{R})$ be a weighted graph with vertices $V$, edges $E$ and $w$ a weight function on $E$. In TSP, the aim is to find the minimum weight loop in $G$ that visits each vertex of $V$ exactly once and returns to the starting point. Let $|V|=n$ and $w_{ij}$ the weight between vertices $i$ and $j$. TSP is then formulated as:

\begin{align}
\min \quad & \sum_{i=1}^{n} \sum_{\substack{j=1 \\ j \neq i}}^{n} w_{ij} \, x_{ij} \label{eq:tsp_objective} \\
\text{s.t.} \quad 
& \sum_{\substack{j=1 \\ j \neq i}}^{n} x_{ij} = 1, && \forall i \label{eq:tsp_constraint_1} \\
& \sum_{\substack{i=1 \\ i \neq j}}^{n} x_{ij} = 1, && \forall j \label{eq:tsp_constraint_2} \\
& \sum_{i \in Q} \sum_{j \in Q} x_{ij} \leq |Q| - 1, && \forall Q \subset V \label{eq:tsp_constraint_3}
\end{align} \label{eq:tsp}

where $x_{ij}=1$ if edge $(i,j)$ is part of the solution, $0$ otherwise. $Q$ is the sub-path with $2 \leq \lvert{Q}\rvert < n-1$. Equality constraints in  \ref{eq:tsp_constraint_1}-\ref{eq:tsp_constraint_2} state that each city can be reached and left only once. On the other side, the inequality constraint in Equation~\ref{eq:tsp_constraint_3} ensures that the solution has a number of edges $\leq |Q|-1$. The Table~\ref{tab:tsp_notation} summarizes the notations used for TSP formulation. 

\begin{table}[h!]
\centering
\caption{Summary of notations and descriptions for the Traveling Salesman Problem (TSP) with one type of example solved as part of the experiments.}
\label{tab:tsp_notation}
\begin{tabular}{|c|c|c|}
\hline
\textbf{Notation} & \textbf{Description} & \textbf{Example} \\ \hline
$V$ & Set of vertices (cities) & -\\ \hline
$E$ & Set of edges between vertices & -\\ \hline
$w: E \mapsto \mathbb{R}$ & Weight function on the edges & 1 \\ \hline
$w_{ij}$ & Weight (distance) between vertex $i$ and vertex $j$ & 1\\ \hline
$G = (V, E, w)$ & A weighted graph with $V$, $E$, and $w$ & -\\ \hline
$n = |V|$ & Total number of vertices (cities) & 3\\ \hline
$x_{ij}$ & \makecell{Binary variable: $1$ if edge $(i, j)$ is part of the solution, \\ $0$ otherwise} & \{0,1\}\\ \hline
$Q$ & Subset of vertices forming a sub-path & -\\ \hline
\end{tabular}
\end{table}

\subsection{Slack Encoding in Constraint Optimization}
Slack encoding is a widely adopted technique for handling inequality constraints in combinatorial optimization in classical computing. It works by introducing additional slack variables to transform inequalities into equalities, which simplifies their integration into penalty-based optimization frameworks. Specifically, for an inequality constraint $h(x_j) \leq 0$, a non-negative slack variable $s_j \geq 0$ is introduced, reformulating the constraint as $h(x_j) + s_j = 0$. This conversion makes it possible to apply standard quadratic penalty terms commonly used in QUBO formulations.

Although slack encoding provides a systematic way to manage inequality constraints, it has two significant drawbacks:
(1) it increases the number of decision variables, thereby directly increasing the required number of qubits,
(2) it uses linear or quadratic penalty terms that may not sufficiently differentiate between small and large constraint violations.

Given these limitations, in the following section, we introduce an alternative approach: an exponential penalty encoding that eliminates the need for extra slack variables while providing a more scalable and qubit-efficient mechanism for penalizing constraint violations.

\section{Exponential Penalization} \label{sec:proposed_method}
Given classical combinatorial optimization problems with constraints, it is necessary to convert them into a Quadratic Unconstrained Binary Optimization (QUBO) formulation to be compatible with quantum computing frameworks. Equality constraints can often be incorporated directly into the objective function as quadratic penalty terms, without introducing additional decision variables. For example, the equality constraints for BPP in Equation~\ref{eqn:bpp_constraint_1}, and for TSP in Equations~\ref{eq:tsp_constraint_1}, \ref{eq:tsp_constraint_2}  can be integrated into the objective functions as follows:

\begin{equation}
    min_{BPP} = \sum_{j=1}^{K} B_j + [\sum_i \sum_{j=1}^{K} x_{ij} - 1]^2
\end{equation}

\begin{equation}
    min_{TSP} = \sum_{i=1}^{n} \sum_{j\neq i, j=1}^{n} w_{ij} x_{ij}+ \sum_j [\sum_{j\neq i, i=1}^{n} x_{ij}-1]^2 + \sum_i [\sum_{j\neq i, j=1}^{n} x_{ij}-1]^2 
\end{equation}

However, applying a similar approach to inequality constraints typically requires transforming these constraints into equalities. This is commonly achieved by introducing additional decision variables, such as slack variables, to bridge the gap between the inequality formulation and the QUBO representation. This process leads to an increase in the problem dimensionality, which in turn increases the number of qubits required to solve the problem on quantum computers. To address this challenge, we propose a novel QUBO encoding inspired by the class of exponential functions. This method integrates inequality constraints directly into the objective function without the need for slack decision variables, thereby preserving the compactness of the problem representation while still ensuring effective constraint enforcement.

\subsection{Class of Exponential Penalty Functions}

We define a general exponential penalty formulation
\begin{equation}
    F_k(x) = f(x) - \sum_{i=1}^m P_k\big(g_i(x)\big),
\end{equation}
where $P_k:\mathbb{R}\to\mathbb{R}$ is a penalty term parameterized by $k$ and $f(x)$ is the objective function corresponding to objective funtion in Equation \ref{eq:cop1}. We study three exponential forms with distinct scaling properties. For $k \in \mathbb{N}$ and constants $a>1$ and $b>a$, define:
\begin{align}
P_k^{(1)}(g_i(x)) &= e^{k g_i(x)},  \label{eq:P1}\\
P_k^{(2)}(g_i(x)) &= \frac{1}{a^k}\, e^{a^k g_i(x)}, \label{eq:P2}\\
P_k^{(3)}(g_i(x)) &= \frac{1}{a^k}\, e^{b^k g_i(x)}. \label{eq:P3}
\end{align}
where $g_i(x)$ is an inequality term \ref{eq:cop_ineq_const}.  The corresponding penalized objectives are $F_r(x) = f(x) - \sum_{i=1}^m P_k^{(r)}(g_i(x))$, for $r\in\{1,2,3\}$.

\subsection{Simple exponential penalty $P^1$}
The penalty in \ref{eq:P1} introduces a direct exponential encoding of constraint violation and is fully governed by scalar parameter $k>0$:
\begin{equation}
F_1(x) = f(x) - \sum_{i=1}^m e^{k g_i(x)},
\end{equation}

The exponential term $e^{k g_i(x)}$ exhibits the following properties:

\begin{enumerate}
    \item \textbf{Curvature:} The first and second derivatives with respect to $g_i(x)$ are
    \[
    \frac{d}{dg_i} e^{k g_i(x)} = k e^{k g_i(x)}, \quad
    \frac{d^2}{dg_i^2} e^{k g_i(x)} = k^2 e^{k g_i(x)}.
    \]
    Both are strictly positive, implying that the penalty is smooth, convex, and strictly increasing in $g_i(x)$.

    \item \textbf{Boundary behavior:} For $g_i(x) \le 0$, the penalty remains bounded above by 1. Near the constraint boundary, a second-order Taylor expansion gives
    \[
    e^{k g_i(x)} = 1 + k g_i(x) + \frac{1}{2}k^2 g_i(x)^2 + O(g_i(x)^3),
    \]
    which shows that local curvature increases quadratically with $k$. Thus, large $k$ values result in rapid growth even for small violations, making this form highly sensitive near the feasible boundary.

    For any fixed violation $g_i(x) > 0$, $e^{k g_i(x)} \to \infty$ as $k \to \infty$. Consequently, infeasible solutions become unbounded in the objective, ensuring convergence toward the feasible region.
\end{enumerate}

\subsection{Scaled exponential penalty $P^2$}
The penalty in \ref{eq:P2} introduces a scale normalization to balance the early- and late-stage effects of the exponential term:
\begin{equation}
F_2(x) = f(x) - \sum_{i=1}^m \frac{1}{a^k} e^{a^k g_i(x)}, \quad a > 1.
\end{equation}

Here, the term $a^{-k}$ counteracts part of the exponential growth, providing a softer penalty when $g_i(x)$ is close to zero, while still ensuring divergence for large violations.

\begin{enumerate}
    \item \textbf{Boundary behavior:} Expanding around $g_i(x) = 0$,
    \[
    \frac{1}{a^k} e^{a^k g_i(x)} = \frac{1}{a^k}\left(1 + a^k g_i(x) + \frac{(a^k g_i(x))^2}{2} + O(g_i(x)^3)\right)
    = \frac{1}{a^k} + g_i(x) + \frac{a^k g_i(x)^2}{2} + O(g_i(x)^3).
    \]
    The linear term is independent of $k$, implying that near the feasible region the penalty responds linearly to small violations, regardless of the penalty strength. This allows gentle exploration around the constraint boundary without excessive gradient magnitudes.

    For $g_i(x) > 0$, the exponential term dominates:
    \[
    \frac{1}{a^k} e^{a^k g_i(x)} \approx a^{-k} e^{a^k g_i(x)} \to \infty \quad \text{as } k \to \infty,
    \]
    so feasibility is still asymptotically enforced.

    \item \textbf{Curvature:} The gradient contribution from the penalty term is \(e^{a^k g_i(x)} \nabla g_i(x)\), which remains modest near feasibility and increases sharply when constraints are violated. The corresponding curvature term scales as \(a^k e^{a^k g_i(x)}\), making the landscape steeper only for substantial violations.
\end{enumerate}

\subsection{Dual-rate exponential panalty $P^3$}
The penalty $P^3$ introduces two scaling parameters, $a$ and $b$, for early-stage smoothness and late-stage enforcement:
\begin{equation}
F_3(x) = f(x) - \sum_{i=1}^m \frac{1}{a^k} e^{b^k g_i(x)}, \quad 1 < a < b.
\end{equation}

This construction generalizes \(F_2\) by allowing independent control over the penalty’s growth rate inside and outside the feasible region.

\begin{enumerate}
    \item \textbf{Boundary behavior:} Near $g_i(x) = 0$,
    \[
    \frac{1}{a^k} e^{b^k g_i(x)} = \frac{1}{a^k}\left(1 + b^k g_i(x) + \frac{1}{2}b^{2k} g_i(x)^2 + O(g_i(x)^3)\right)
    = \frac{1}{a^k} + \frac{b^k}{a^k} g_i(x) + \frac{b^{2k}}{2a^k} g_i(x)^2 + O(g_i(x)^3).
    \]
    The linear term grows as $(b/a)^k$, so increasing $b/a$ intensifies the gradient pressure near feasibility, allowing flexible adjustment of local sensitivity.

    For $g_i(x) > 0$,
    \[
    \frac{1}{a^k} e^{b^k g_i(x)} \to \infty,
    \]
    at a rate exponentially faster than $F_2$, since $b^k > a^k$. Hence, constraint violations are penalized more aggressively, ensuring faster convergence toward feasibility in later stages.

    \item \textbf{Curvature:} The gradient contribution is $(b^k / a^k) e^{b^k g_i(x)} \nabla g_i(x)$, and the curvature scales as $(b^{2k}/a^k) e^{b^k g_i(x)}$. This formulation allows one to control both the magnitude and sharpness of the penalty through the ratio $b/a$.
\end{enumerate}

These exponential penalty functions has the property that they are uniformly bounded over any compact feasible region, and again, by an appropriate choice of parameters, the gradient remains uniformly bounded over the feasible region. 

By comparing the general form of exponential function and slack approaches empirically, we derived the following statements that show the relation between the number of variables in the problem and required qubits: 
\begin{boxA}
\begin{stmt}\label{stmt:1}
     Exponential penalization approach requires $N\cdot K+K$ qubits for BPP instance with $N$ items and $K$ bins. 
\end{stmt}
\end{boxA}

\begin{boxA}
\begin{stmt}\label{stmt:2}
    Exponential penalty approach requires $n\cdot (n-1)$ qubits for TSP instances with $n$ vertices.
\end{stmt}   
\end{boxA}


\section{Experimental Setup} \label{sec:experiment}
This section details the experimental framework employed to assess the performance of the proposed exponential penalty encoding method. We describe the experimental setup, evaluation metrics, hyperparameters, and dataset selection.

The \textit{exponential penalization} encoding is applied to BPP and TSP inequality constraints and converted into QUBO formulation. Then we solved the resulting QUBO using the Quantum Approximate Optimization Algorithm (QAOA). The QAOA was first introduced in~\cite{farhi2014quantum} as a VQA to find approximate solutions to optimization problems. The key idea behind QAOA is to encode the objective function of the optimization problem into a Hamiltonian to search for an optimal bitstring that will give an optimal solution with high probability. We select the most commonly used classical optimizer \textit{Constrained Optimization by Linear Approximation (COBYLA)}~\cite{powell1994direct} as it proved to be the fastest among other competitors such as \textit{Simultaneous Perturbation Stochastic Approximation (SPSA)}~\cite{spall1987stochastic}, Adam and AMSGRAD optimizers~\cite{kingma2014adam}. COBYLA is a gradient-free optimization method that uses linear approximations to optimize the objective function and is particularly useful for problems where the gradient of the objective function is not known or difficult to calculate. We executed all instances on local Aer simulator \textit{ibmq\_qasm\_simulator} provided by Qiskit \footnote{https://www.ibm.com/quantum/qiskit}. This simulator can simulate both ideal and added noisy quantum circuits up to 32 qubits.

\subsection{Evaluation Metrics}
The performance of the proposed method and results from QAOA are evaluated by the following metrics.
\paragraph{Qubit Reduction:}  
We evaluate how the number of required qubits \( Q(n) \) grows with problem size \( n \). To quantify improvements over baseline approach $Q_{\text{slack}}$, we define the qubit reduction ratio as:
\begin{equation}
    Q_{RE} = 1-\frac{Q_{\text{exp}}(n)}{Q_{\text{slack}}(n)}    
\end{equation}
where $Q_{slack}$ and $Q_{exp}$ are the number of qubits required for slack and exponential encoding respectively. 
A higher ratio indicates greater qubit efficiency, demonstrating the proposed method's suitability for NISQ-era devices. The higher $Q_{RE}$ is, the less information is needed to find the solution for the same problem. 

\paragraph{Solution Quality:}  
We evaluate the solution accuracy by comparing the objective function values obtained from the quantum and classical algorithms. This is quantified using the Mean Squared Error (MSE) across $k$ problem instances:
\begin{equation}
    \text{MSE} = \frac{1}{k} \sum_{i=1}^{k} (C_i - Q_i)^2
\end{equation}
where $ C_i $ and $ Q_i $ represent the classical and quantum objective values for the $ i^{\text{th}} $ instance, respectively.

\paragraph{Convergence Time:} Another metric is the computational time spent to solve the instances of BPP and TSP. Multiple factors affect the execution time, i.e., the number of decision variables per instance, the classical optimizer, and the number of iterations. For comparative analysis, we also consider the relative time efficiency:

\begin{equation}
    Q_T = \frac{T_{\text{slack}}}{T_{\text{exp}}}
\end{equation}

\subsection{Hyperparameter Tuning}
We conduct a hyperparameter optimization to find the best parameters for our encoding. Table~\ref{tab:encoding_parameters} shows the parameters we used in our experiment. For the choice of $k$, we considered a uniform distribution within a bounded range $[0, N]$, with $N \in Z$. The choice for $r,s$ can be seen in Table \ref{tab:encoding_parameters}. 
\begin{table}[h!] 
\centering
\caption{Different examples from the class of exponential penalization encoding with corresponding parameters. For all simulations, additive penalty parameters $p=[1,10]$ are considered. }
\label{tab:encoding_parameters}
\begin{tabular}{|c|c|c|c|}
\hline
Problem Type      & F1($r=k, s=1$) & F2($r=a^k, s=a^k$) & F3($r=a^k, s=b^k$) \\ \hline
BPP/TSP   &   $k = [0,..,10]$ & $k = [0,..,10]$, $a=[2,3,4]$ & $k = [0,..,10]$, $\{a,b\}=[2,3,4]$  \\ \hline
\end{tabular}
\end{table}

\subsection{Dataset and Experimental Testbed}
We generated synthetic instances of BPP and TSP considering the quantum computer limits such as execution time and availability of qubits. A set of samples for each problem type was generated to check the overall quality of the proposed encoding method. 

For the implementation and analysis, we use Python 3.11 and Qiskit on a Lenovo ThinkPad X1 Yoga Intel(R) Core(TM) i7-8550U CPU @ 1.80GHz. The experiments were performed on server AMD EPYC 7452 32-Core Processor with 128 CPUs, 2 sockets, 32 cores per socket, and 2 threads per core. 

\section{Results and Discussion} \label{sec:results}
\subsection{Qubit reduction}
\begin{figure}[t!]
    \caption{Number of Qubits w.r.t. Slack and Decision Variables Required to Encode problems. }
    \label{fig:qubit_variable_relation}
    \begin{subfigure}{0.4\textwidth}
    \centering
        \includegraphics[width=\textwidth]{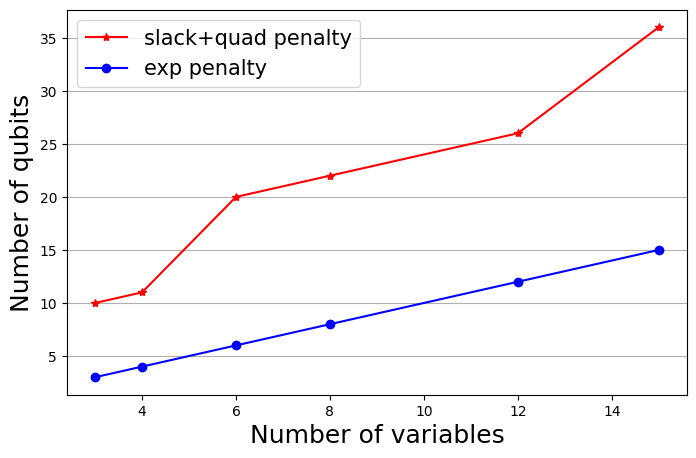}
        \caption{BPP}
    \end{subfigure}
    \hfill
    \begin{subfigure}{0.4\textwidth}
    \centering
        \includegraphics[width=\textwidth]{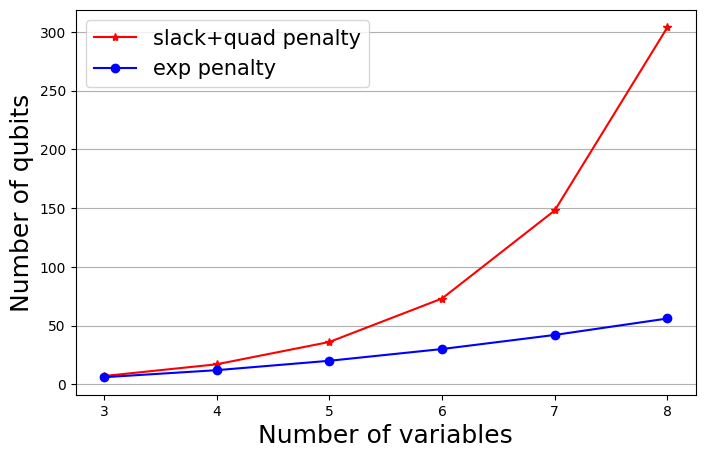}
        \caption{TSP }
    \end{subfigure}

\end{figure}

We empirically evaluate $Q_{RE}$ for BPP and TSP, using exponential penalization and slack penalties using statements~\ref{stmt:1} and~\ref{stmt:2}. In Figure~\ref{fig:qubit_variable_relation}, we see a qubit count reduction of up to $57$\% for BPP and $83\%$ for TSP when applying exponential penalties compared to the slack approach. Moreover, the number of qubits is increasing linearly for BPP and exponentially for TSP with the increase in the number of decision variables. The results of our method are shown using a blue line. The red line demonstrates the nonlinear increase of qubits with the number of variables when applying the slack method. This is because of the additional slack variables used to convert inequality to equality, which causes the number of qubits to grow faster. Table~\ref{tab:comparison} summarizes the number of qubits for BPP and TSP using slack, followed by quadratic penalties and exponential penalty method as stated in Statement~\ref{stmt:1} and~\ref{stmt:2}.

\begin{table}[!h]
\setlength{\tabcolsep}{10pt} 
\renewcommand{\arraystretch}{1.5} 
    \centering
    \caption{Comparison of number of qubits using slack+quadratic penalty and exponential penalty 
    for BPP and TSP. By applying the exponential penalty function, the number of qubits required by the encoding is reduced by $O(2^n)$ and $O(K)$ for TSP and BPP, respectively. }
    \begin{tabular}{|@{ }c@{ }|@{ }c@{ }|@{ }c@{ }|}
        \hline
         Problem & Slack+Quadratic & \textbf{Exponential (our approach)} \\
         \hline
         BPP &  $N\cdot K+K+O(K)$ & $\mathbf{N\cdot K+K}$\\
         TSP & $n^2+O(2^n)$ & $\mathbf{n^2-n}$\\
         \hline
    \end{tabular}
    
    \label{tab:comparison}
\end{table}

\subsection{Solution Quality}
\begin{figure}[h!]
        \caption{QAOA objective function (energy values) for (a) BPP and (b) TSP with respect to varying $\beta, \gamma$ values on noiseless simulation.}
        \begin{subfigure}{0.5\textwidth}
        \centering
            \includegraphics[width=1\textwidth]{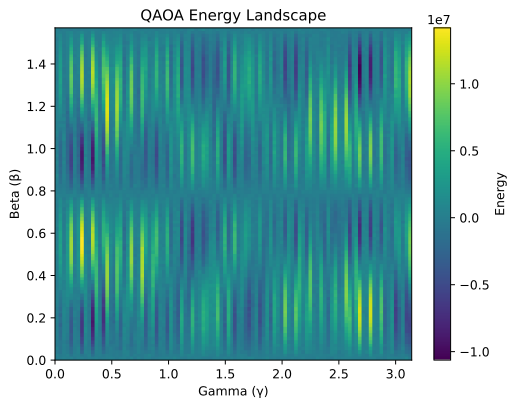}
            \caption{BPP with $(r,s,p)=(17,7,8)$}
            \label{fig:bpp_qaoa_landscape_exp}
        \end{subfigure}
        \hfill
        \begin{subfigure}{0.5\textwidth}
        \centering
            \includegraphics[width=1\textwidth]{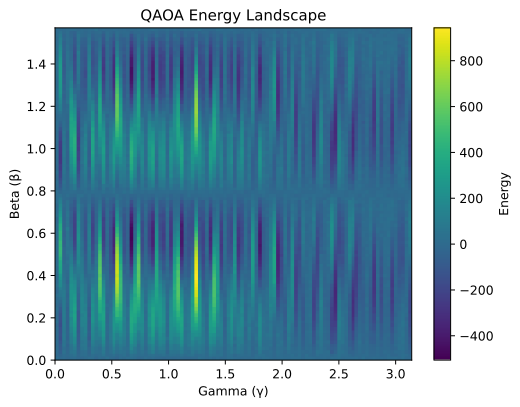}
            \caption{TSP with $(r,s,p)=(10,5,1)$}
            \label{fig:tsp_qaoa_landscape_exp}
        \end{subfigure}

\end{figure}

QAOA objective function value with respect to $\beta$ and $\gamma$ values for $p=1$ with noiseless simulation is in Figures~\ref{fig:bpp_qaoa_landscape_exp}-\ref{fig:tsp_qaoa_landscape_exp} for BPP and TSP, respectively.

\subsection{Approximation Probability}
In variational quantum algorithms such as QAOA, for each optimization iteration, the quantum circuit is executed multiple times (shots), and each execution yields a bitstring that corresponds to a solution. 
The frequency with which each bitstring appears reflects its approximation probability. For instance, in the last classical optimization iteration step, if a specific bitstring appears with high frequency, it is considered the most likely solution produced by the algorithm. Our experimental results indicate that classical and quantum solutions (from QAOA) are similar when it comes to bitstring,
which means we achieve a correct solution by QAOA. 
For BPP and TSP, we utilized 8 and 12 qubits, respectively, resulting in $2^8$ and $2^{12}$ possible solution samples. Under the assumption of a uniform probability distribution, the likelihood of obtaining any specific solution is $\frac{1}{2^8} = 0.04\%$ for BPP and $\frac{1}{2^{12}} = 0.0002\%$ for TSP. Notably, our encoding approach significantly improves the approximation probability of the optimal solution, achieving over 6\% for BPP and over 21\% for TSP, as shown in Table~\ref{tab:approximation_probability}. 

\begin{wraptable}{r}{0.5\textwidth}
\centering
\caption{The approximation probability for a single example from each problem type with the exponential penalization method. QAOA here is executed on noiseless simulator with $p=1$ layer. And values represent the percentage.}
\label{tab:approximation_probability}
\begin{tabular}{|c|ccc|}
\hline
\multirow{2}{*}{Problem type}   & \multicolumn{3}{c|}{QAOA}                        \\ \cline{2-4} 
                                & \multicolumn{1}{c|}{F1}    & \multicolumn{1}{c|}{F2}          & \multicolumn{1}{c|}{F3}        \\ \hline 
BPP                             & \multicolumn{1}{c|}{4.72\%}      & \multicolumn{1}{c|}{5.03\%}            & \multicolumn{1}{c|}{6.35\%}                \\
TSP                             & \multicolumn{1}{c|}{8.52\%}      & \multicolumn{1}{c|}{8.25\%}            & \multicolumn{1}{c|}{21.26\%}                \\
\hline
\end{tabular}
\end{wraptable}

\begin{figure}[h!]
        \caption{The mean convergence of QAOA with exponential encoding for single problem samples executed on noiseless simulators with hyperparameters $r=10, s=5, p=2$.}
        \label{fig:mean_convergence}
        \begin{subfigure}{0.5\textwidth}
        \centering
            \includegraphics[width=\textwidth]{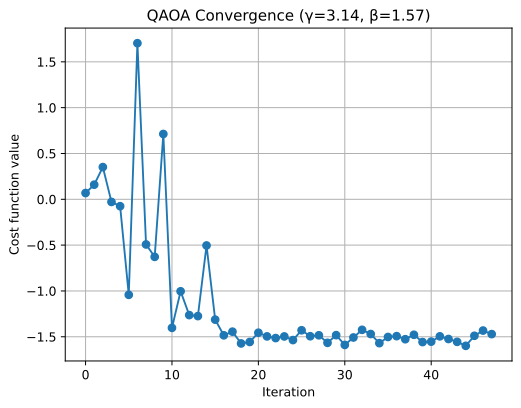}
            \caption{BPP}
            \label{fig:bpp_mean_convergence}
        \end{subfigure}
        \hfill
        \begin{subfigure}{0.5\textwidth}
        \centering
            \includegraphics[width=1\textwidth]{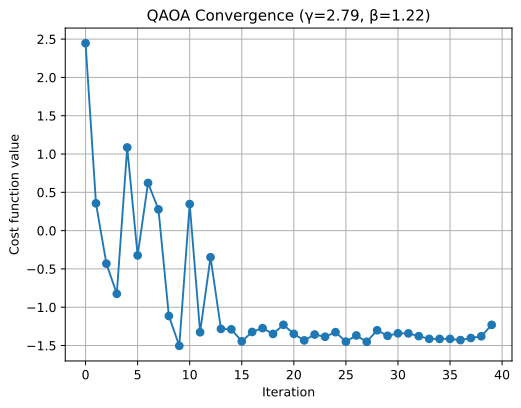}
            \caption{TSP}
            \label{fig:tsp_mean_convergence}
        \end{subfigure}
    
\end{figure}

Across all tested instances, the vast majority of samples converged to the correct solution. Only 0.2\% of BPP instances and 5\% of TSP instances do not converge to optimal solution, indicating that the correlation between high approximation probability and low objective value is consistently strong. Additionally, the mean convergence of QAOA with exponential encoding can be seen in Figures ~\ref{fig:bpp_mean_convergence} and ~\ref{fig:tsp_mean_convergence}.

\subsection{Convergence Time}
Convergence time for BPP with 8 qubits and TSP with 12 qubits instances are, on average, 7 and 23 seconds, respectively. While the slack variable approach requires a considerably higher amount of qubits, we do not consider the slack variable approach in the convergence time analysis. QAOA execution of large-scale samples that exceed 20 qubits takes too long to run. For example, BPP sample with bins 2, items 3 requires 8 qubits with exponential encoding, while with slack encoding, it requires 24 qubits. Relatively, simulating the QAOA solution of such an instance on a noise-free quantum computer takes around 5 seconds with exponential encoding and over 6 hours with slack encoding, as it requires 24 qubits.

\subsection{Hyperparameter tuning}
Our exponential encoding method relies on three key hyperparameters: ${r, s, p}$. As a pre-processing step, we optimized these parameters individually for each problem type, exploring the search space defined by $r \in [1,20]$, $s \in [1,20]$, and $p \in [1,10]$. Empirical results demonstrate that, for BPP, the selection of these hyperparameters has a substantial influence on the quality of the resulting solutions. The optimization process was carried out using OPTUNA, a state-of-the-art hyperparameter optimization framework~\cite{optuna_2019}. 




\section{Conclusion and Future Works} \label{sec:conclusion}
We proposed a generalized exponential penalization framework for encoding inequality constraints into QUBO formulations for combinatorial optimization problems. Our approach eliminates the need for slack variables, resulting in a significant reduction in qubit requirements up to 57\% for BPP and 83\% for the TSP. Experimental evaluations using QAOA demonstrate that the exponential encoding not only achieves competitive solution quality but also improves the approximation probability of optimal solutions, with success rates exceeding 6\% for BPP and 21\% for TSP. 

Future research will focus on extending this framework to larger problem instances, exploring adaptive parameter tuning, and validating its performance on real quantum hardware.




\bibliographystyle{unsrt}
\bibliography{references}

\begin{thebibliography}{10}

\bibitem{hervert2020production}
Laura Hervert-Escobar and Jesus~Fabi{\'a}n L{\'o}pez-P{\'e}rez.
\newblock Production planning and scheduling optimization model: a case of study for a glass container company.
\newblock {\em Annals of Operations Research}, 286:529--543, 2020.

\bibitem{qi2020optimization}
Chengming Qi and Lishuan Hu.
\newblock Optimization of vehicle routing problem for emergency cold chain logistics based on minimum loss.
\newblock {\em Physical Communication}, 40:101085, 2020.

\bibitem{gupta2022enhanced}
Neha Gupta, Kamali Gupta, Deepali Gupta, Sapna Juneja, Hamza Turabieh, Gaurav Dhiman, Sandeep Kautish, and Wattana Viriyasitavat.
\newblock Enhanced virtualization-based dynamic bin-packing optimized energy management solution for heterogeneous clouds.
\newblock {\em Mathematical Problems in Engineering}, 2022:1--11, 2022.

\bibitem{sankar2024benchmarking}
Krishanu Sankar, Artur Scherer, Satoshi Kako, Sam Reifenstein, Navid Ghadermarzy, Willem~B Krayenhoff, Yoshitaka Inui, Edwin Ng, Tatsuhiro Onodera, Pooya Ronagh, et~al.
\newblock A benchmarking study of quantum algorithms for combinatorial optimization.
\newblock {\em npj Quantum Information}, 10(1):64, 2024.

\bibitem{abbas2024challenges}
Amira Abbas, Andris Ambainis, Brandon Augustino, Andreas B{\"a}rtschi, Harry Buhrman, Carleton Coffrin, Giorgio Cortiana, Vedran Dunjko, Daniel~J Egger, Bruce~G Elmegreen, et~al.
\newblock Challenges and opportunities in quantum optimization.
\newblock {\em Nature Reviews Physics}, pages 1--18, 2024.

\bibitem{glover2022quantum}
Fred Glover, Gary Kochenberger, Rick Hennig, and Yu~Du.
\newblock Quantum bridge analytics i: a tutorial on formulating and using qubo models.
\newblock {\em Annals of Operations Research}, 314(1):141--183, 2022.

\bibitem{dantzig1990origins}
George~B Dantzig.
\newblock Origins of the simplex method.
\newblock In {\em A history of scientific computing}, pages 141--151. 1990.

\bibitem{mirkarimi2024quantum}
Puya Mirkarimi, Ishaan Shukla, David~C Hoyle, Ross Williams, and Nicholas Chancellor.
\newblock Quantum optimization with linear ising penalty functions for customer data science.
\newblock {\em Physical Review Research}, 6(4):043241, 2024.

\bibitem{montanez2024unbalanced}
JA~Montanez-Barrera, Dennis Willsch, Alberto Maldonado-Romo, and Kristel Michielsen.
\newblock Unbalanced penalization: A new approach to encode inequality constraints of combinatorial problems for quantum optimization algorithms.
\newblock {\em Quantum Science and Technology}, 9(2):025022, 2024.

\bibitem{coello2022constraint}
Carlos A~Coello Coello.
\newblock Constraint-handling techniques used with evolutionary algorithms.
\newblock In {\em Proceedings of the genetic and evolutionary computation conference companion}, pages 1310--1333, 2022.

\bibitem{Tan2021qubitefficient}
Benjamin Tan, Marc-Antoine Lemonde, Supanut Thanasilp, Jirawat Tangpanitanon, and Dimitris~G. Angelakis.
\newblock Qubit-efficient encoding schemes for binary optimisation problems.
\newblock {\em {Quantum}}, 5:454, May 2021.

\bibitem{10646502}
Manuel Schnaus, Lilly Palackal, Benedikt Poggel, Xiomara Runge, Hans Ehm, Jeanette~Miriam Lorenz, and Christian~B. Mendl.
\newblock { Efficient Encodings of the Travelling Salesperson Problem for Variational Quantum Algorithms }.
\newblock In {\em 2024 IEEE International Conference on Quantum Software (QSW)}, pages 81--87, Los Alamitos, CA, USA, July 2024. IEEE Computer Society.

\bibitem{leonidas2024qubit}
Ioannis~D Leonidas, Alexander Dukakis, Benjamin Tan, and Dimitris~G Angelakis.
\newblock Qubit efficient quantum algorithms for the vehicle routing problem on noisy intermediate-scale quantum processors.
\newblock {\em Advanced Quantum Technologies}, 7(5):2300309, 2024.

\bibitem{roch2023effect}
Christoph Roch, Daniel Ratke, Jonas N{\"u}{\ss}lein, Thomas Gabor, and Sebastian Feld.
\newblock The effect of penalty factors of constrained hamiltonians on the eigenspectrum in quantum annealing.
\newblock {\em ACM Transactions on Quantum Computing}, 4(2):1--18, 2023.

\bibitem{v2023hybrid}
Sebasti{\'a}n V.~Romero, Eneko Osaba, Esther Villar-Rodriguez, Izaskun Oregi, and Yue Ban.
\newblock Hybrid approach for solving real-world bin packing problem instances using quantum annealers.
\newblock {\em Scientific Reports}, 13(1):11777, 2023.

\bibitem{MANDAL199891}
C.A. Mandal, P.P. Chakrabarti, and S.~Ghose.
\newblock Complexity of fragmentable object bin packing and an application.
\newblock {\em Computers and Mathematics with Applications}, 35(11):91--97, 1998.

\bibitem{farhi2014quantum}
Edward Farhi, Jeffrey Goldstone, and Sam Gutmann.
\newblock A quantum approximate optimization algorithm.
\newblock {\em arXiv preprint arXiv:1411.4028}, 2014.

\bibitem{powell1994direct}
Michael~JD Powell.
\newblock {\em A direct search optimization method that models the objective and constraint functions by linear interpolation}.
\newblock Springer, 1994.

\bibitem{spall1987stochastic}
James~C Spall.
\newblock A stochastic approximation technique for generating maximum likelihood parameter estimates.
\newblock In {\em 1987 American control conference}, pages 1161--1167. IEEE, 1987.

\bibitem{kingma2014adam}
Diederik~P Kingma and Jimmy Ba.
\newblock Adam: A method for stochastic optimization.
\newblock {\em arXiv preprint arXiv:1412.6980}, 2014.

\bibitem{optuna_2019}
Takuya Akiba, Shotaro Sano, Toshihiko Yanase, Takeru Ohta, and Masanori Koyama.
\newblock Optuna: A next-generation hyperparameter optimization framework.
\newblock In {\em Proceedings of the 25th {ACM} {SIGKDD} International Conference on Knowledge Discovery and Data Mining}, 2019.

\end{thebibliography}

\end{document}